\documentclass[aip,
rsi,%
 amsmath,amssymb,
 reprint,%
]{revtex4-1}

\usepackage{graphicx}
\usepackage{dcolumn}
\usepackage{bm}
\usepackage{subfigure}
\usepackage[usenames]{color}

\begin{document}

\preprint{AIP/123-QED}

\title{Comparing Theory and Simulation for Thermo-osmosis}

\author{Karel Proesmans}
\email{Karel.Proesmans@uhasselt.be}
 \affiliation{Hasselt University, B-3590 Diepenbeek, Belgium.}
\author{Daan Frenkel}%
\affiliation{ 
Department of Chemistry, University of Cambridge, Lensfield Road, Cambridge CB2 1EW, UK
}%

\date{\today}

\begin{abstract}
We report a numerical study of thermo-osmotic slip, i.e.~the particle flux induced by a thermal gradient along a solid-fluid interface. To facilitate comparison with theory, we consider a model of an ideal but viscous gas. We compare three numerical routes to obtain the slip coefficient: 1.~by using the Onsager reciprocity relations 2.~by using the appropriate Green-Kubo relation 3.~via the excess enthalpy. The numerical results are found to be mutually consistent, and to agree with the theoretical prediction based on the assumption that hydrodynamics and thermodynamics are locally valid. 

\end{abstract}
\keywords{thermo-osmotic slip, linear irreversible thermodynamics}
\maketitle

\section{Introduction}
With the increasing importance of nano-scale transport, both in man-made devices (nano-fluidics) and in cell biology, there is a great need to improve our ``microscopic'' understanding of phoretic transport, i.e. transport that occurs only in the presence of interfaces. 

Whilst there exists an extensive literature on this subject~\cite{anderson1989colloid,levich1962physicochemical,derjaguin1987surface,de2013non,wurger2010thermal}, the focus of earlier publications has been on local continuum descriptions. Subsequently, there have been several publications on mesoscopic~\cite{yang2013thermophoretically,lusebrink2012thermophoresis} and atomistic~\cite{han2005thermophoresis,ganti2017molecular,ganti2018hamiltonian} simulations of phoretic flows, but there is a shortage of simulations on model systems that are simple enough to allow quantitative comparison  between analytical theory and simulation. 

In this paper, we present such a study, by considering  thermo-osmotic flow of a simple model gas, described by Multi-Particle Collision (MPC) dynamics (also known as Stochastic Rotation Dynamics (SRD)) \cite{gompper2009multi}. The MPC fluid obeys the equation of state of an ideal gas, but it undergoes collisions that mediate the transport of particles, energy and momentum. The transport properties of the MPC model are (to an excellent approximation) known analytically~\cite{gompper2009multi}.  As a consequence, we can not only compute the thermo-osmotic slip coefficients for this fluid numerically but we can also predict them analytically. 

We stress that the interest of the present paper does not lie in the computed numbers as such (after all, the model that we used was chosen for simplicity, not for realism), but simply in the fact that all different methods that we used to compute thermo-osmotic slip appear to be consistent. Hence, future authors may select any of the methods used, based on considerations of convenience. 

In section \ref{lit}, we give a brief overview of the relevant framework of linear irreversible thermodynamics. This framework will then be used in section \ref{tm} to obtain an expression for the thermo-osmotic slip coefficient of our model system. In section \ref{nr}, we will compare these theoretical predictions with numerical numerical results. 

\section{Linear irreversible thermodynamics\label{lit}}
In reviewing the framework of linear irreversible thermodynamics in the context of thermo-osmotic slip, we focus  on those aspects that are most directly relevant (we will, for instance, ignore magnetic fields and rotating systems). For a more thorough introduction, we refer to the standard text by De Groot and Mazur~\cite{de2013non}.

A system can be driven out of equilibrium by a {thermodynamic force}, $\mathcal{F}_i$. This driving force could be a temperature gradient, a chemical-potential gradient or a mechanical force. Every thermodynamic force induces a {thermodynamic flux}, $J_i$, such as a heat flux or a particle flux. The entropy production rate, $\sigma$, is then given by
\begin{equation}
    \sigma=\sum_i \mathcal{F}_iJ_i.
\end{equation}

The central assumption of linear irreversible thermodynamics is that the thermodynamic forces are small enough to justify a first-order Taylor expansion of the thermodynamic fluxes in terms of the thermodynamic forces,
\begin{equation}
    J_i=\sum_j L_{ij}\mathcal{F}_j,\label{Jdef}
\end{equation}
with
\begin{equation}
    L_{ij}=\left.\frac{\partial J_i}{\partial \mathcal{F}_j}\right|_{\textrm{eq}},
\end{equation}
the Onsager coefficients. Here, the subscript {\em eq} stands for the equilibrium condition, $\mathcal{F}_i=0$ for all $i$. The above expansion is typically valid if the relation between forces and fluxes is analytical. 
The Onsager coefficients obey an important symmetry, namely the Onsager reciprocal relations,
\begin{eqnarray}
    L_{ij}=L_{ji}.
\end{eqnarray}
This property has several important consequences. For example, it implies that one can determine an Onsager coefficient, even without directly measuring the associated flux.

The Onsager coefficients can also be determined from the equilibrium correlations of the system,
\begin{equation}
    L_{ij}=\frac{1}{k_B}\int^{\infty}_0dt\,\left\langle J_i(0)J_j(t)\right\rangle_\textrm{eq},\label{gk}
\end{equation}
where $k_B$ is the Boltzmann constant. This relation, known as the Green-Kubo relation, gives another method to determine the Onsager coefficients without directly applying a thermodynamic force.

\section{Model system\label{tm}}
\begin{figure*}[htp]
  \centering
  \subfigure[Schematic setup of the system.\label{fig1a}]{\includegraphics[scale=0.35]{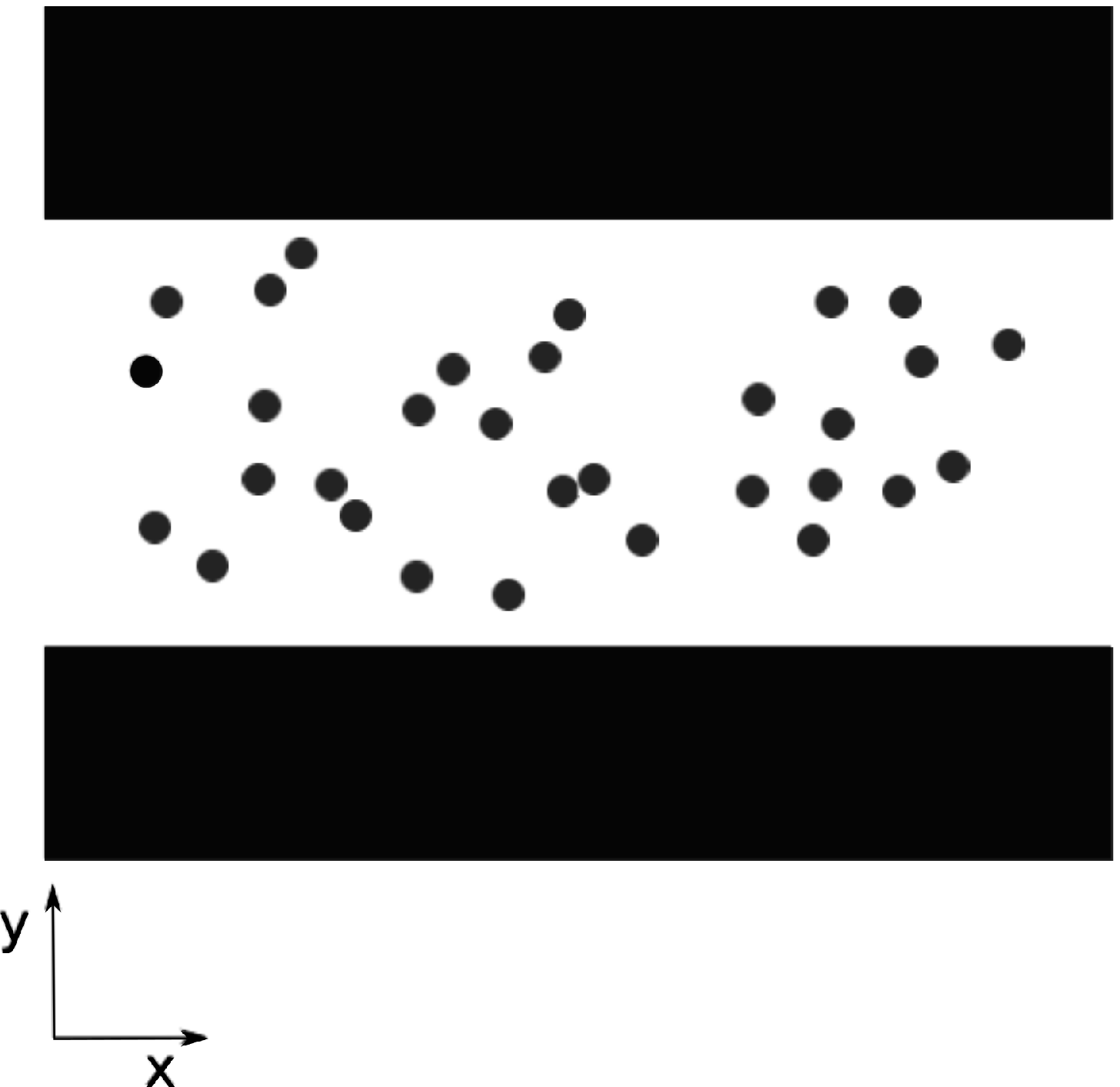}}\qquad\quad\qquad
  \subfigure[Cross-section of the potential $U(y)$.\label{fig1b}]{\includegraphics[scale=0.5]{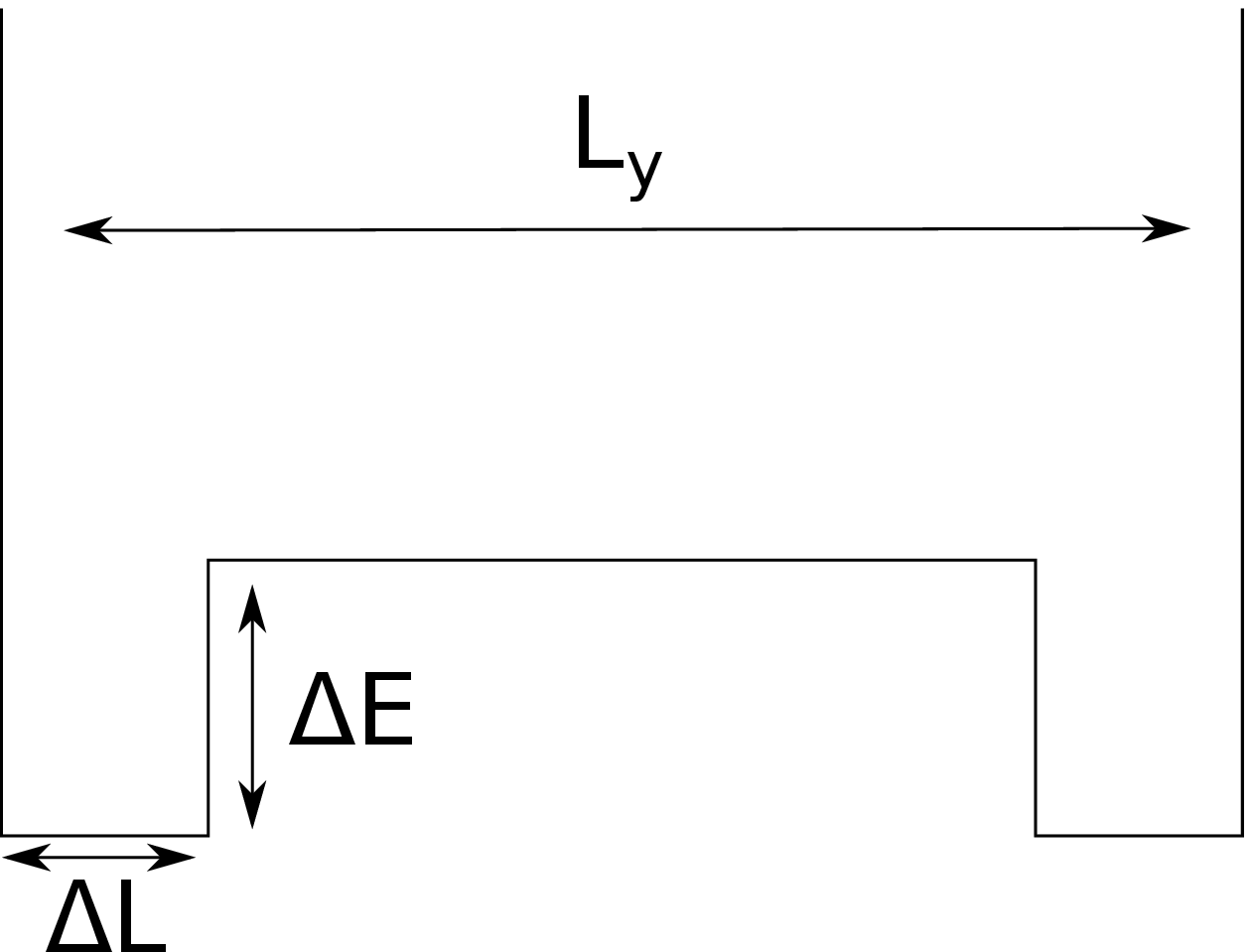}}
  \caption{Setup of the system \label{fig1}}
\end{figure*}
In this section we describe the MPC model that we used to study  thermo-osmotic slip. 
This model was chosen for simplicity. In particular, our aim was to consider a model where simulations can be compared directly with theoretical predictions. 
\subsection{Model}
The model system that we studied was  a two-dimensional MPC gas confined between two parallel flat walls, separated by a distance $L_y$ (see Fig.~\ref{fig1}.)  We used periodic boundary conditions along the slit direction. In our simulations, we impose no-slip boundary conditions at the hard walls. Away from the no-slip planes, the walls interact with the gas particles via a square-well potential of width $\Delta L$ and depth $\Delta E$.   In the spirit of the MPC model, we assume that the gas particles do not interact with each other directly,  i.e.~they behave like an ideal gas. However, the MPC particles may thermalize via collisions. 

In our simulations we impose  a thermal gradient $\nabla T$ and a body force inducing a pressure gradient $\nabla P$ along the slit direction (the $x$-axis). These gradients result in  heat and particle fluxes, $Q_x$ and $v_x$. The entropy production rate can be written as \cite{ganti2017molecular}
\begin{equation}
    \sigma=-\left(\overline{Q_x}-h_B \overline{v_x}\,\overline{\rho}\right)\frac{\nabla T}{T^2}-\overline{v_x}\frac{\nabla P}{T},
\end{equation}
where $\sigma$ and $\overline{Q_x}$ are the entropy production and heat flux per unit of volume, $\overline{v_x}$ is the average velocity per particle, ${h}_B$ is the average enthalpy per particle in the bulk (i.e.~away from the walls), $\overline{\rho}$ is the average density of the gas, $\nabla P$ is the pressure gradient along the $x$-direction, and $T$ is the temperature of the system.
For small gradients the fluxes can be written as (Eq.~(\ref{Jdef})):
\begin{eqnarray}
    \overline{Q_x}-h_B \overline{v_x}\,\overline{\rho}&=&-L_{11} \frac{\nabla T}{T^2}-L_{12}\frac{\nabla P}{T}\\
    \overline{v_x}&=&-L_{21} \frac{\nabla T}{T^2}-L_{22}\frac{\nabla P}{T}.
\end{eqnarray}
The off-diagonal Onsager coefficient $L_{21}$ now determines the thermo-osmotic slip, as it gives the particle flux induced by a temperature gradient. 

We can now determine explicit expressions for the fluxes $\overline{v_x}$ and $\overline{Q_x}-h_B\overline{v_x}\,\overline{\rho}$. The average velocity is by definition given by
\begin{equation}
    \overline{v_x}=\frac{\sum_{i=1}^N v_{x;i}}{N}=\frac{\int^{L_y}_0dy\, \rho(y)v_x(y)}{L_y\overline{\rho}},\label{vdef}
\end{equation}
where $ \rho(y)v_x(y)$ is defined as $\sum_i \delta(y_i-y)v_{x,i}$.
The heat flux can be written as,
\begin{eqnarray}
    \overline{Q_x}&=&\frac{\sum_i\left(\frac{m\left(v_{x;i}^2+v_{y;i}^2\right)}{2}+u(y_i)\right)v_{x;i}}{L_xL_y}
\end{eqnarray}
with potential energy $u(y)$ given by
\begin{equation}
    u(y)=\begin{cases}
    -\Delta E & y<\Delta L\\
    0 & \Delta L<y<L_y-\Delta L\\
    -\Delta E & y<L_y-\Delta L
    \end{cases},
\end{equation}
cf.~Fig.~\ref{fig1b}.
This potential mimics the interaction of the particles with the wall via a potential well. 
To make further progress, we split the velocity in a thermal part, $v_{x/y;\textrm{th}}$ and a drift part $v_{x;\textrm{drift}}$ (only present in the x-direction), which we assume to be independent. Furthermore, we assume that $\overline{ v_{x/y;\textrm{th}}}=\overline{ v_{x/y;\textrm{th}}^3}=0$ and $\overline{ v_{x;\textrm{drift}}^2}\ll \overline{ v_{x/y;\textrm{th}}^2}=k_BT/(2m)$, where the last equality follows from the equipartition theorem. One can now write
\begin{equation}
    \overline{v_{x}^3}=\overline{\left(v_{x;\textrm{th}}+v_{x;\textrm{drift}}\right)^3}\approx \frac{3}{2}k_BT \overline{v_{x}},\quad  \overline{v_{y}^2v_x}=\frac{k_BT\overline{v_{x}}}{2}.
\end{equation}
which leads to a simplification for the heat flux,
\begin{equation}    \overline{Q_x}\approx \overline{\rho}\left({2k_BT\overline{v_{x}}+\overline{u(y)v_x}}\right).
\end{equation}
Invoking the equipartition theorem once more leads to a simple formula for the enthalpy of particles in the bulk,
\begin{equation}
    h_B=\frac{\sum_i \frac{m\left(v_{x;i}^2+v_{y;i}^2\right)}{2}+PV_{\textrm{B}}}{N_{\textrm{B}}}={2k_BT},
\end{equation}
where the sum is over all particles in the bulk. Therefore, the total flux is given by
\begin{equation}
    \overline{Q_x}-h_B \overline{v_x}\,
    \overline{\rho}=\frac{\sum_{i}u(y_i)v_{x;i}}{L_xL_y}=\frac{\int_0^{L_y}\rho(y)u(y)v_x(y)}{L_y},\label{qdef}
\end{equation}
where $ \rho(y)u(y)v_x(y)$ is defined as $\sum_i \delta(y_i-y)u(y_i)v_{x,i}$.

\subsection{Navier-Stokes equation}
To determine more explicit forms for the fluxes, we first assume that the flow should satisfy a (linearized) Navier-Stokes equation,
\begin{equation}
   \frac{d}{dy}\left(\eta(y)\frac{d}{dy}v_x(y)\right)=\nabla P(y),\label{NS}
\end{equation}
where $\eta(y)$ is the viscosity. 
Furthermore, we assume that the thermodynamic forces are small enough for a local equilibrium ansatz to hold:
\begin{eqnarray}
    \rho(x,y)&=&\begin{cases}
     \rho_{\textrm{W}}& y<\Delta L \\
    \rho_{\textrm{B}}  & \Delta L<y<L_y-\Delta L\\
 \rho_{\textrm{W}}   & y>L_y-\Delta L \\
    \end{cases}.
\end{eqnarray}
with
\begin{eqnarray}
    \rho_{\textrm{W}}&=&\frac{\overline{\rho} L_y e^{\beta\Delta E}}{L_y-2\Delta L+2\Delta Le^{\beta\Delta E}}\\
    \rho_{\textrm{B}}&=&\frac{\overline{\rho}L_y}{L_y-2\Delta L+2\Delta Le^{\beta\Delta E}}.
\end{eqnarray}
It should be stressed that, although independent of $y$, $\rho_{\textrm{W}}$ and $ \rho_{\textrm{B}}$ can vary along the $x$-axis, either via the density or via a temperature gradient.
For the MCP fluid, the viscosity only depends on 
the density~\cite{gompper2009multi} and hence it should also be piecewise constant, $\eta(y)=\eta_W$ if $y<\Delta L$ or $y>L_y-\Delta L$ and $\eta(y)=\eta_B$ otherwise.

In the situations that we will consider, the pressure gradient is either induced by a constant body force or by a temperature gradient, therefore it will also be piecewise constant along the $y$-axis ,
\begin{eqnarray}
    \nabla P(x,y)&=&\begin{cases}
     \nabla P_{\textrm{W}}& y<\Delta L \\
    \nabla P_{\textrm{B}}  & \Delta L<y<L_y-\Delta L\\
 \nabla P_{\textrm{W}}   & y>L_y-\Delta L \\
    \end{cases},
\end{eqnarray}
where the specific values of $\nabla P_{\textrm{B}}$ and $\nabla P_{\textrm{W}}$ depend on the driving forces.

Under the assumption of no-slip boundary condition and using the fact that the velocity profile should be continuous, one can now solve the Navier-Stokes equation, Eq.~(\ref{NS}):
\begin{equation}
    v_x(y)=\begin{cases}
  v_{x;\textrm{W}}(y)& y<\Delta L,\\
   v_{x;\textrm{B}}(y)& \Delta L\leq y \leq L_y-\Delta L\\
    v_{x;\textrm{W}}(L_y-y)& y>L_y-\Delta L
  \end{cases}\label{vprof}
\end{equation}
with
\begin{eqnarray}
    v_{x;\textrm{W}}(y)&=&\frac{\nabla P_{\textrm{W}}y^2-\left(\left(L_y-2\Delta L\right)\nabla P_{\textrm{B}}+2\Delta L\nabla P_{\textrm{W}}\right)y}{2\eta_W}\nonumber\\\\
     v_{x;\textrm{B}}(y)&=&\frac{\left(\Delta L(L_y-\Delta L)-y(L_y-y)\right)\nabla P_{\textrm{B}}}{2\eta_B}\nonumber\\&&-\frac{\Delta L(L_y-2\Delta L)\nabla P_{\textrm{B}}+\Delta L^2\nabla P_{\textrm{W}}}{2\eta_W}
\end{eqnarray}

One can now use this result to determine the thermodynamic fluxes in terms of these quantities. The velocity flux can be derived from Eq.~(\ref{vdef}),
\begin{widetext}
\begin{multline}
   \overline{v_x}=-\frac{\nabla P_{\textrm{B}}\rho_B}{12\eta_B L_y\overline{\rho}}\left(L_y-2\Delta L\right)^3-\frac{2\nabla P_{\textrm{W}}\rho_W\Delta L^3}{3\eta_W L_y\overline{\rho}}-\frac{\rho_{\textrm{B}}}{2\eta_W L_y\overline{\rho}}\left(L_y-2\Delta L\right)\left(\Delta L(L_y-2\Delta L)\nabla P_{\textrm{B}}+\Delta L^2\nabla P_{\textrm{W}}\right)\\-\frac{\rho_W}{2\eta_W\overline{\rho}L_y}\Delta L^2(L_y-2\Delta L)\nabla P_{\textrm{B}}\label{vav}
\end{multline}
\end{widetext}
while the heat flux, Eq.~(\ref{qdef}), is given by
\begin{multline}
    \overline{Q_x}-h_B\overline{v_x} \,\overline{\rho}=\frac{\Delta E\Delta L^2\rho_W}{6\eta_W L_y}\times\\\left(3(L_y-2\Delta L)\nabla P_{\textrm{B}}+4\Delta L \nabla P_{\textrm{W}}\right)
\end{multline}
This completes our general analysis of the thermodynamic fluxes of a gas in a tube, as we have determined the two independent fluxes purely in terms of external parameters. To make further progress, one needs to specify the driving. This will be done in the next section to determine the Onsager coefficients.

\subsection{Onsager coefficients}
With these ingredients, we are now ready to determine the off-diagonal Onsager coefficients. We first calculate the heat flux induced by a pressure gradient. As the force on every particle is equal, the pressure gradients near the wall and in the bulk are given by
\begin{eqnarray}
    \nabla P_{\textrm{W}}&=&\frac{\nabla P L_ye^{\beta\Delta E}}{L_y-2\Delta L+2\Delta Le^{\beta\Delta E}}\\
     \nabla P_{\textrm{B}}&=&\frac{\nabla P L_y}{L_y-2\Delta L+2\Delta Le^{\beta\Delta E}},
\end{eqnarray}
respectively. This leads to an Onsager coefficient,
\begin{equation}
    L_{12}=-\frac{T\overline{\rho}\Delta E L_y\Delta L^2e^{\beta\Delta E}\left(3(L_y-2\Delta L)+4\Delta Le^{\beta\Delta E}\right)}{6\eta_W\left(L_y-2\Delta L+2\Delta L e^{\beta \Delta E}\right)^2}.
\end{equation}

On the other hand, one can try to derive the transport coefficient associated the thermo-osmotic slip directly, by studying the situation in the presence of a temperature gradient but in the absence of a bulk pressure gradient,
\begin{equation}
    \nabla P_{\textrm{B}}=k_B\left(\rho_{\textrm{B}} \nabla T+T\nabla \rho_{\textrm{B}}\right)=0,\label{pwdb}
\end{equation}
or
\begin{equation}
    \nabla \rho_{\textrm{B}}=-\frac{\rho_{\textrm{B}} \nabla T}{T}.
\end{equation}
The gradient of the density near the wall can be calculated as
\begin{eqnarray}
\nabla \rho_{\textrm{W}}&=&\nabla\left(e^{\frac{\Delta E}{k_BT}}\rho_{\textrm{B}}\right)\nonumber\\
&=&-\frac{\left(\frac{\Delta E}{k_BT}+1\right)}{T}\rho_{\textrm{W}}\nabla T
\end{eqnarray}
\begin{eqnarray}
    \nabla P_{\textrm{W}}&=&k_B\left(T\nabla \rho_{\textrm{W}}+\rho_{\textrm{W}}\nabla T\right)\nonumber\\
    &=&-\frac{\Delta E\rho_{\textrm{W}}\nabla T}{T}\label{pwdt}
\end{eqnarray}
and therefore, Eq.~(\ref{vav}) simplifies to
\begin{equation}
    \overline{v_x}=\frac{2\Delta E\rho_{\textrm{W}}^2\Delta L^2\nabla T}{T\eta_W\overline{\rho}}\left(\frac{\Delta L}{3}+\frac{e^{-\beta\Delta E}(L_y-2\Delta L)}{4}\right),
\end{equation}
which determines the thermo-osmotic slip coefficient
\begin{equation}L_{21}=-\frac{T\overline{\rho}\Delta EL_y\Delta L^2e^{\beta\Delta E}\left(3(L_y-2\Delta L)+4\Delta Le^{\beta\Delta E}\right)}{6\eta_W\left(L_y-2\Delta L+2\Delta L e^{\beta \Delta E}\right)^2}.\end{equation}
One immediately verifies that the expressions for the off-diagonal Onsager coefficients $L_{12}$ and $L_{21}$ are equivalent and therefore, Onsager symmetry is indeed valid.

In conclusion, we have determined the transport coefficient associated with the thermo-osmotic slip fully in terms of predetermined variables. Furthermore, one can see from the above calculation that the thermo-osmotic slip can be determined even in the absence of a temperature gradient due to Onsager symmetry.

\subsection{Comparison with Derjaguin's method}
The standard method to determine the thermo-osmotic slip in a fluid is by using Dejaguin's method, which states that the slip induced by a single wall is determined by the excess enthalpy, $\Delta h(y)=h(y)-h_\textrm{B}$
\begin{equation}L_{21}^{\textrm{Der}}=\frac{T}{2}\int^\infty_0dy\, \frac{y\rho(y)\Delta h(y)}{\eta(y)}.\label{Der}\end{equation}
In the system under study we have two walls with limited interaction range, leading to a Derjaguin Onsager coefficient equal to \cite{derjaguin1987surface,ganti2017molecular}
\begin{eqnarray}
    L_{21}^{\textrm{Der}}&=&T\int^{\frac{L_y}{2}}_0dy\,\frac{y\rho(y)\Delta h(y)}{\eta(y)} \nonumber\\
    &=&-\frac{T\Delta EL_y\Delta L^2e^{\beta\Delta E}\overline{\rho}}{2\eta_W\left(L_y-2\Delta L+2\Delta L e^{\beta\Delta E}\right)}.
\end{eqnarray}

This raises the question whether our result is compatible with Derjaguin's method. 
One can verify that this is indeed the case if one takes into account the extra assumption $\Delta L\ll L_y$.

\section{Numerical results \label{nr}}
With an explicit prediction for the thermo-osmotic slip at hand, we are now ready to test the theory with numerical simulations. For realistic hydrodynamic simulations, we rely on a stochastic rotation dynamics (SRD) algorithm. We shall first discuss the implementation of the SRD algorithm. Subsequently, we will use the algorithm to determine the thermo-osmotic slip coefficient $L_{21}$.

\subsection{Stochastic rotation dynamics}
In this algorithm, the tube of particles is partitioned into square simulation cells with length $a$. At every time-step, all particles in a cell get a new velocity, $\mathbf{v}_f$ given by
\begin{equation}
    \mathbf{v}_f=\mathbf{v}_{\textrm{cell}}+\mathbf{R}\left(\mathbf{v}_i-\mathbf{v}_{\textrm{cell}}\right)
\end{equation}
with $\mathbf{v}_{\textrm{cell}}$ the average velocity of the particles inside the simulation cell, $\mathbf{v}_i$ the velocity of the particle before the SRD step, and $\mathbf{R}$ a rotation matrix,
\begin{equation}
    \mathbf{R}=\left( \begin{array}{cc}
      \cos \alpha  &-\sin \alpha  \\
       \sin \alpha  & \cos\alpha
    \end{array}\right),
\end{equation}
 $\alpha$ being a rotation angle which is fixed during a simulation. Furthermore, there are also particles in the wall with fixed temperature. This leads to a thermalization of gas particles with the wall. For a throughout review of the SRD algorithm, we refer to the literature\cite{gompper2009multi}.

To predict a value for the thermo-osmotic transport coefficient, one needs to determine the friction coefficient $\eta$. This coefficient is known for an SRD fluid \cite{ihle2005equilibrium,tuzel2003transport},
\begin{multline}
    \eta=\frac{k_BT\Delta t\rho}{2m}\left(\frac{\rho a^2}{\left(\rho a^2-1+e^{-\rho a^2}\right)\sin^2 \alpha}-1\right)\\+\frac{\rho a^2-1+e^{-\rho a^2}}{12\Delta t}(1-\cos(\alpha)),
\end{multline}
where $\Delta t$ is the duration of one SRD simulation step.

All simulations will be done with $\alpha=0.722\pi$, $a=1$ and $\Delta t=0.1$. Furthermore, we shall take periodic boundary conditions in the x-direction, and set $L_x=10$, $L_y=50$, $N=5000$ $\Delta L=10$, $\Delta E=4$, $k_BT=10$ and $m=1$. The theoretically predicted value for the thermo-osmotic transport coefficient is then given by
\begin{equation}
    L_{21}^{\textrm{Theory}}=-9.90\cdot 10^2.\label{Ltheo}
\end{equation}

\subsection{Results}
\begin{figure}[htp]
  \centering
 \includegraphics[scale=0.42]{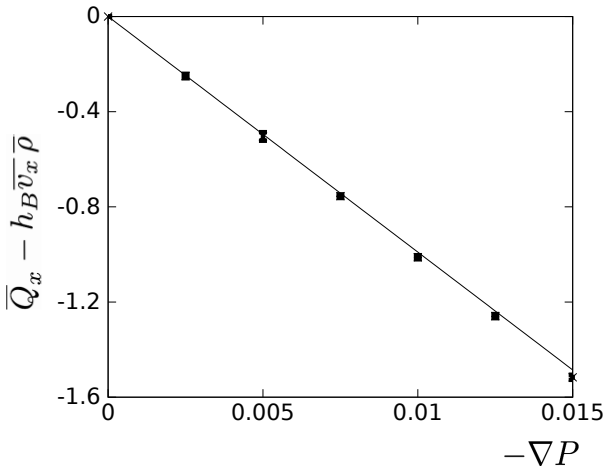}
  \caption{Numerically measurement of heat flux $\overline{Q_x}-h_B\overline{v_x}\,\overline{\rho}$ (dots), with theoretically predicted curve (full line).}
  \label{figons}
\end{figure}
To determine the thermo-osmotic transport coefficient, we explicitly use Onsager symmetry, by applying a body force to the fluid particles and calculating the induced heat flux. This gives the thermo-osmotic transport coefficient,
\begin{equation}
    L_{21}=L_{12}=\frac{\left\langle Q_x-\bar{h}_{\textrm{B}} v_x\right\rangle}{-\nabla P/T},
\end{equation}
which should hold for small pressure gradients. In agreement with our theoretical analysis, we apply an equal force $f$ in the positive $x$ direction to every particle, which induces a pressure gradient $\nabla P=-f\overline{\rho}$. We vary $f$ between $0$ and $1.5\cdot 10^{-3}$ or $-\nabla P=1.5\cdot 10^{-2}$, cf.~Fig.~\ref{figons}. After averaging over $20$ runs of duration $t=4\cdot 10^5$, we get
\begin{equation}
    L_{21}^{\textrm{Ons}}=-\left(1.007\pm 0.005\right)\cdot 10^3,
\end{equation}
which is in good agreement with the theoretically predicted value. The small discrepancy between the two results can be explained from the approximations that we made in determining the theoretical value of the transport coefficient.

\begin{figure}[htp]
  \centering
 \includegraphics[scale=0.42]{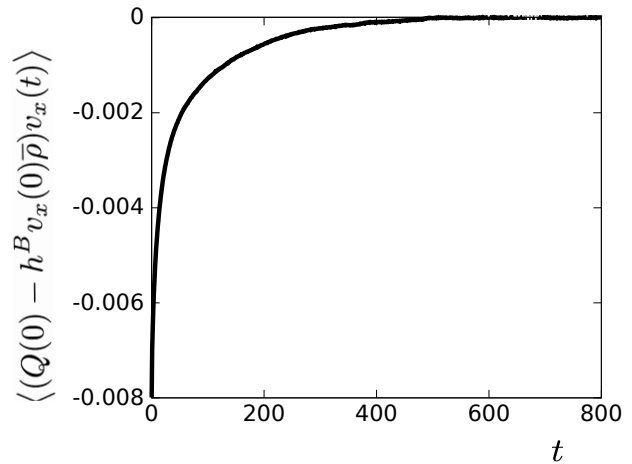}
  \caption{Numerically measurement of correlation function $\left\langle (Q(0)-h^Bv_x(0)\overline{\rho})v_x(t)\right\rangle$.}
  \label{gkfig}
\end{figure}
An alternative numerical route to obtain the thermo-osmotic coefficient is via the Green-Kubo relation, Eq.~(\ref{gk}). As the infinite integration cannot be done analytically, we need to introduce a cut-off time for the simulation. From Fig.~\ref{gkfig}, one can observe that $\langle (Q(0)-h^Bv_x(0))v_x(t)\rangle\approx 0$ if $t\gtrsim 600$. Therefore, we shall calculate
\begin{equation}
    L^{\textrm{GK}}_{21}=L_xL_y\int^{600}_0dt\left\langle (Q(0)-h^Bv_x(0)\overline{\rho})v_x(t)\right\rangle.
\end{equation}
Averaging over $1.6\cdot 10^5$ runs gives
\begin{equation}
    L^{\textrm{GK}}_{21}=-\left(1.05\pm 0.05\right)\cdot 10^3,
\end{equation}
which again agrees reasonably well with the theoretical predictions, although the precision is one order of magnitude worse than the Onsager method.

Whenever a temperature gradient is applied over the tube, the particles near the wall feel an effective body force related to the excess enthalpy \cite{ganti2017molecular},
\begin{equation}
    f(y)=-\frac{\left(h(y)-h^B\right)\nabla T}{T}.\label{fh}
\end{equation}
Note that this is indeed in agreement with Eqs.~(\ref{pwdb}) and (\ref{pwdt}).
Explicitly applying this force (in the absence of a real temperature gradient) to the system gives a third computational method to calculate the thermo-osmotic slip coefficient. If we set $-\nabla T=0.00125$ in Eq.~(\ref{fh}), and let the system run $20$ times for a duration of $t=4\cdot 10^5$
we get
\begin{equation}
    L^{\textrm{h}}_{21}=\frac{\overline{v_x}}{-\nabla T/T^2}=\left(9.9\pm 0.2\right)\cdot 10^3.
\end{equation}
This again is in agreement with the theoretical prediction.

\section{Conclusion\label{con}}
In this work, we have studied thermo-osmosis in a simple model consisting of an ideal gas. In particular, we have compared a theoretical approach based on hydrodynamic arguments with computational approaches based on Onsager symmetry, on the Green-Kubo relation and on the excess enthalpy. The resulting thermo-osmotic slip coefficients are in good agreement with each other, implying that mesoscopic simulations are indeed consistent with local continuum descriptions.

\begin{acknowledgments}
KP is a postdoctoral fellow of the Research Foundation-Flanders (FWO). The computational resources and services used in this work were provided by the VSC (Flemish Supercomputer Center), funded by the Research Foundation - Flanders (FWO) and the Flemish Government department EWI
\end{acknowledgments}

\providecommand{\noopsort}[1]{}\providecommand{\singleletter}[1]{#1}%


\begin{thebibliography}{13}%
\makeatletter
\providecommand \@ifxundefined [1]{%
 \@ifx{#1\undefined}
}%
\providecommand \@ifnum [1]{%
 \ifnum #1\expandafter \@firstoftwo
 \else \expandafter \@secondoftwo
 \fi
}%
\providecommand \@ifx [1]{%
 \ifx #1\expandafter \@firstoftwo
 \else \expandafter \@secondoftwo
 \fi
}%
\providecommand \natexlab [1]{#1}%
\providecommand \enquote  [1]{``#1''}%
\providecommand \bibnamefont  [1]{#1}%
\providecommand \bibfnamefont [1]{#1}%
\providecommand \citenamefont [1]{#1}%
\providecommand \href@noop [0]{\@secondoftwo}%
\providecommand \href [0]{\begingroup \@sanitize@url \@href}%
\providecommand \@href[1]{\@@startlink{#1}\@@href}%
\providecommand \@@href[1]{\endgroup#1\@@endlink}%
\providecommand \@sanitize@url [0]{\catcode `\\12\catcode `\$12\catcode
  `\&12\catcode `\#12\catcode `\^12\catcode `\_12\catcode `\%12\relax}%
\providecommand \@@startlink[1]{}%
\providecommand \@@endlink[0]{}%
\providecommand \url  [0]{\begingroup\@sanitize@url \@url }%
\providecommand \@url [1]{\endgroup\@href {#1}{\urlprefix }}%
\providecommand \urlprefix  [0]{URL }%
\providecommand \Eprint [0]{\href }%
\providecommand \doibase [0]{http://dx.doi.org/}%
\providecommand \selectlanguage [0]{\@gobble}%
\providecommand \bibinfo  [0]{\@secondoftwo}%
\providecommand \bibfield  [0]{\@secondoftwo}%
\providecommand \translation [1]{[#1]}%
\providecommand \BibitemOpen [0]{}%
\providecommand \bibitemStop [0]{}%
\providecommand \bibitemNoStop [0]{.\EOS\space}%
\providecommand \EOS [0]{\spacefactor3000\relax}%
\providecommand \BibitemShut  [1]{\csname bibitem#1\endcsname}%
\let\auto@bib@innerbib\@empty
\bibitem [{\citenamefont {Anderson}(1989)}]{anderson1989colloid}%
  \BibitemOpen
  \bibfield  {author} {\bibinfo {author} {\bibfnamefont {J.~L.}\ \bibnamefont
  {Anderson}},\ }\href@noop {} {\bibfield  {journal} {\bibinfo  {journal}
  {Annual review of fluid mechanics}\ }\textbf {\bibinfo {volume} {21}},\
  \bibinfo {pages} {61} (\bibinfo {year} {1989})}\BibitemShut {NoStop}%
\bibitem [{\citenamefont {Levich}(1962)}]{levich1962physicochemical}%
  \BibitemOpen
  \bibfield  {author} {\bibinfo {author} {\bibfnamefont {V.~G.}\ \bibnamefont
  {Levich}},\ }\href@noop {} {\  (\bibinfo {year} {1962})}\BibitemShut
  {NoStop}%
\bibitem [{\citenamefont {Derjaguin}\ \emph {et~al.}(1987)\citenamefont
  {Derjaguin}, \citenamefont {Churaev}, \citenamefont {Muller},\ and\
  \citenamefont {Kisin}}]{derjaguin1987surface}%
  \BibitemOpen
  \bibfield  {author} {\bibinfo {author} {\bibfnamefont {B.~V.}\ \bibnamefont
  {Derjaguin}}, \bibinfo {author} {\bibfnamefont {N.~V.}\ \bibnamefont
  {Churaev}}, \bibinfo {author} {\bibfnamefont {V.~M.}\ \bibnamefont {Muller}},
  \ and\ \bibinfo {author} {\bibfnamefont {V.}~\bibnamefont {Kisin}},\
  }\href@noop {} {\emph {\bibinfo {title} {Surface forces}}}\ (\bibinfo
  {publisher} {Springer},\ \bibinfo {year} {1987})\BibitemShut {NoStop}%
\bibitem [{\citenamefont {De~Groot}\ and\ \citenamefont
  {Mazur}(2013)}]{de2013non}%
  \BibitemOpen
  \bibfield  {author} {\bibinfo {author} {\bibfnamefont {S.~R.}\ \bibnamefont
  {De~Groot}}\ and\ \bibinfo {author} {\bibfnamefont {P.}~\bibnamefont
  {Mazur}},\ }\href@noop {} {\emph {\bibinfo {title} {Non-equilibrium
  thermodynamics}}}\ (\bibinfo  {publisher} {Courier Corporation},\ \bibinfo
  {year} {2013})\BibitemShut {NoStop}%
\bibitem [{\citenamefont {W{\"u}rger}(2010)}]{wurger2010thermal}%
  \BibitemOpen
  \bibfield  {author} {\bibinfo {author} {\bibfnamefont {A.}~\bibnamefont
  {W{\"u}rger}},\ }\href@noop {} {\bibfield  {journal} {\bibinfo  {journal}
  {Reports on Progress in Physics}\ }\textbf {\bibinfo {volume} {73}},\
  \bibinfo {pages} {126601} (\bibinfo {year} {2010})}\BibitemShut {NoStop}%
\bibitem [{\citenamefont {Yang}\ and\ \citenamefont
  {Ripoll}(2013)}]{yang2013thermophoretically}%
  \BibitemOpen
  \bibfield  {author} {\bibinfo {author} {\bibfnamefont {M.}~\bibnamefont
  {Yang}}\ and\ \bibinfo {author} {\bibfnamefont {M.}~\bibnamefont {Ripoll}},\
  }\href@noop {} {\bibfield  {journal} {\bibinfo  {journal} {Soft Matter}\
  }\textbf {\bibinfo {volume} {9}},\ \bibinfo {pages} {4661} (\bibinfo {year}
  {2013})}\BibitemShut {NoStop}%
\bibitem [{\citenamefont {L{\"u}sebrink}, \citenamefont {Yang},\ and\
  \citenamefont {Ripoll}(2012)}]{lusebrink2012thermophoresis}%
  \BibitemOpen
  \bibfield  {author} {\bibinfo {author} {\bibfnamefont {D.}~\bibnamefont
  {L{\"u}sebrink}}, \bibinfo {author} {\bibfnamefont {M.}~\bibnamefont {Yang}},
  \ and\ \bibinfo {author} {\bibfnamefont {M.}~\bibnamefont {Ripoll}},\
  }\href@noop {} {\bibfield  {journal} {\bibinfo  {journal} {Journal of
  Physics: Condensed Matter}\ }\textbf {\bibinfo {volume} {24}},\ \bibinfo
  {pages} {284132} (\bibinfo {year} {2012})}\BibitemShut {NoStop}%
\bibitem [{\citenamefont {Han}(2005)}]{han2005thermophoresis}%
  \BibitemOpen
  \bibfield  {author} {\bibinfo {author} {\bibfnamefont {M.}~\bibnamefont
  {Han}},\ }\href@noop {} {\bibfield  {journal} {\bibinfo  {journal} {Journal
  of colloid and interface science}\ }\textbf {\bibinfo {volume} {284}},\
  \bibinfo {pages} {339} (\bibinfo {year} {2005})}\BibitemShut {NoStop}%
\bibitem [{\citenamefont {Ganti}, \citenamefont {Liu},\ and\ \citenamefont
  {Frenkel}(2017)}]{ganti2017molecular}%
  \BibitemOpen
  \bibfield  {author} {\bibinfo {author} {\bibfnamefont {R.}~\bibnamefont
  {Ganti}}, \bibinfo {author} {\bibfnamefont {Y.}~\bibnamefont {Liu}}, \ and\
  \bibinfo {author} {\bibfnamefont {D.}~\bibnamefont {Frenkel}},\ }\href@noop
  {} {\bibfield  {journal} {\bibinfo  {journal} {Physical review letters}\
  }\textbf {\bibinfo {volume} {119}},\ \bibinfo {pages} {038002} (\bibinfo
  {year} {2017})}\BibitemShut {NoStop}%
\bibitem [{\citenamefont {Ganti}, \citenamefont {Liu},\ and\ \citenamefont
  {Frenkel}(2018)}]{ganti2018hamiltonian}%
  \BibitemOpen
  \bibfield  {author} {\bibinfo {author} {\bibfnamefont {R.}~\bibnamefont
  {Ganti}}, \bibinfo {author} {\bibfnamefont {Y.}~\bibnamefont {Liu}}, \ and\
  \bibinfo {author} {\bibfnamefont {D.}~\bibnamefont {Frenkel}},\ }\href@noop
  {} {\bibfield  {journal} {\bibinfo  {journal} {Physical review letters}\
  }\textbf {\bibinfo {volume} {121}},\ \bibinfo {pages} {068002} (\bibinfo
  {year} {2018})}\BibitemShut {NoStop}%
\bibitem [{\citenamefont {Gompper}\ \emph {et~al.}(2009)\citenamefont
  {Gompper}, \citenamefont {Ihle}, \citenamefont {Kroll},\ and\ \citenamefont
  {Winkler}}]{gompper2009multi}%
  \BibitemOpen
  \bibfield  {author} {\bibinfo {author} {\bibfnamefont {G.}~\bibnamefont
  {Gompper}}, \bibinfo {author} {\bibfnamefont {T.}~\bibnamefont {Ihle}},
  \bibinfo {author} {\bibfnamefont {D.}~\bibnamefont {Kroll}}, \ and\ \bibinfo
  {author} {\bibfnamefont {R.}~\bibnamefont {Winkler}},\ }in\ \href@noop {}
  {\emph {\bibinfo {booktitle} {Advanced computer simulation approaches for
  soft matter sciences III}}}\ (\bibinfo  {publisher} {Springer},\ \bibinfo
  {year} {2009})\ pp.\ \bibinfo {pages} {1--87}\BibitemShut {NoStop}%
\bibitem [{\citenamefont {Ihle}, \citenamefont {T{\"u}zel},\ and\ \citenamefont
  {Kroll}(2005)}]{ihle2005equilibrium}%
  \BibitemOpen
  \bibfield  {author} {\bibinfo {author} {\bibfnamefont {T.}~\bibnamefont
  {Ihle}}, \bibinfo {author} {\bibfnamefont {E.}~\bibnamefont {T{\"u}zel}}, \
  and\ \bibinfo {author} {\bibfnamefont {D.~M.}\ \bibnamefont {Kroll}},\
  }\href@noop {} {\bibfield  {journal} {\bibinfo  {journal} {Physical Review
  E}\ }\textbf {\bibinfo {volume} {72}},\ \bibinfo {pages} {046707} (\bibinfo
  {year} {2005})}\BibitemShut {NoStop}%
\bibitem [{\citenamefont {T{\"u}zel}\ \emph {et~al.}(2003)\citenamefont
  {T{\"u}zel}, \citenamefont {Strauss}, \citenamefont {Ihle},\ and\
  \citenamefont {Kroll}}]{tuzel2003transport}%
  \BibitemOpen
  \bibfield  {author} {\bibinfo {author} {\bibfnamefont {E.}~\bibnamefont
  {T{\"u}zel}}, \bibinfo {author} {\bibfnamefont {M.}~\bibnamefont {Strauss}},
  \bibinfo {author} {\bibfnamefont {T.}~\bibnamefont {Ihle}}, \ and\ \bibinfo
  {author} {\bibfnamefont {D.~M.}\ \bibnamefont {Kroll}},\ }\href@noop {}
  {\bibfield  {journal} {\bibinfo  {journal} {Physical Review E}\ }\textbf
  {\bibinfo {volume} {68}},\ \bibinfo {pages} {036701} (\bibinfo {year}
  {2003})}\BibitemShut {NoStop}%
\end{thebibliography}
\end{document}